\documentclass[sigplan]{acmart}

\AtBeginDocument{%
  \providecommand\BibTeX{{%
    \normalfont B\kern-0.5em{\scshape i\kern-0.25em b}\kern-0.8em\TeX}}}

\copyrightyear{2020} 
\acmYear{2020} 
\setcopyright{acmcopyright}
\acmConference[LCTES '20]{Proceedings of the 21st ACM SIGPLAN/SIGBED Conference on Languages, Compilers, and Tools for Embedded Systems}{June 16, 2020}{London, United Kingdom}
\acmBooktitle{Proceedings of the 21st ACM SIGPLAN/SIGBED Conference on Languages, Compilers, and Tools for Embedded Systems (LCTES '20), June 16, 2020, London, United Kingdom}
\acmPrice{15.00}

\usepackage[linesnumbered,ruled]{algorithm2e}
\usepackage{threeparttable}
\newtheorem{Tlemma}{Lemma}
\newenvironment{Lemma}[1]
{\begin{Tlemma}\noindent\textsc{}\itshape}
{\end{Tlemma}}
\newtheorem{Tdef}{Definition}
\newenvironment{Definition}[1]
    {\begin{Tdef}\noindent\textsc{(#1)}\itshape}
    {\end{Tdef}}
\newcommand{\actor}[1]{\textbf{\emph{{#1}}}}
\newcommand{\ineq}[1]{\footnotesize$#1$\normalsize}{}
\usepackage{multirow}
\newcommand{\mc}[1]{\textcolor{black}{#1}}

\SetCommentSty{mycommfont}    

\begin{document}
\fancyhead{}
\title{Compiling Spiking Neural Networks to Neuromorphic Hardware}

\author{Shihao Song, Adarsha Balaji, Anup Das, Nagarajan Kandasamy, and James Shackleford}
\email{{shihao.song, ab3586, anup.das, nk78, jas64}@drexel.edu}
\affiliation{%
  \institution{Electrical and Computer Engineering, Drexel University}
  \streetaddress{P.O. Box 1212}
  \city{Philadelphia}
  \state{PA}
  \postcode{19104}
}








\renewcommand{\shortauthors}{Song, et al.}

\begin{abstract}
 Machine learning applications that are implemented with spike-based computation model, e.g., Spiking Neural Network (SNN), have a great potential to lower the energy consumption when executed on a neuromorphic hardware. However, compiling and mapping an SNN to the hardware is challenging, especially when compute and storage resources of the hardware (viz. crossbars) need to be shared among the neurons and synapses of the SNN.
We propose an approach to analyze and compile 
SNNs on resource-constrained neuromorphic hardware, 
providing
guarantees on key performance metrics such as execution time and throughput. Our approach makes the following three key contributions. First, we propose a greedy technique to partition an SNN into clusters of neurons and synapses such that each cluster can fit on to the resources of a crossbar.
Second, we exploit the rich semantics and expressiveness of Synchronous Dataflow Graphs (SDFGs) 
to represent 
a clustered SNN 
and analyze its performance using Max-Plus Algebra, 
considering
the available compute and storage capacities, buffer sizes, and communication bandwidth. 
Third, 
we propose a self-timed execution-based fast technique to
compile and admit
SNN-based applications to a neuromorphic hardware at run-time, 
adapting dynamically to the available resources on the hardware.
We evaluate our approach with standard SNN-based applications and demonstrate a significant performance improvement compared to 
current practices.
\end{abstract}




\maketitle

\section{Introduction}
\label{sec:introduction}
Machine learning tasks implemented with spike model~\cite{brette2015philosophy} and brain-inspired learning algorithms~\cite{dan2004spike}, e.g., Spiking Neural Network (SNN) \cite{maass1997networks}, 
have a great potential to lower the energy consumption when they are executed on a neuromorphic hardware such as DYNAP-SE~\cite{Moradi_etal18}, TrueNorth~\cite{debole2019truenorth}, Neurogrid~\cite{benjamin2014neurogrid}, SpiNNaker~\cite{furber2014spinnaker}, and Loihi~\cite{davies2018loihi}.
This makes SNNs attractive for implementing machine learning applications in resource and power-constrained environments, ones where sensor and edge devices of the Internet-of-Things (IoT)~\cite{gubbi2013internet} typically operate.
\mc{
A neuromorphic hardware consists of computation units called crossbars, communicating with each other using an interconnect. 
A crossbar can accommodate a fixed number of neurons and synapses.
}

Executing a program on a hardware involves several steps: compilation, resource allocation, and run-time mapping. Although apparent for mainstream computers, these steps are challenging and not very well defined when executing an SNN-based machine learning application on a neuromorphic hardware. This is because 
a neuromorphic hardware implements accumulation-based alternate computing, where neural computations and synaptic storage are co-located inside each crossbar and distributed in the hardware.
This 
is different from a conventional computer where CPUs compute by exchanging data centrally from the memory.

Prior research efforts such as \cite{balaji20pycarl,balaji2019mapping,ji2016neutrams,balaji2019CALframework,das2018mapping,ji2018bridge,lee2019system} have \textit{only} addressed design-time analysis of an application with \textit{unlimited} hardware resources, e.g., arbitrarily large crossbars and many interconnected crossbars as needed to accommodate all neurons and synapses of the application.
While these efforts are still relevant when designing the hardware, they cannot provide a realistic guarantee of performance when executing these applications on an off-the-shelf neuromorphic hardware. This is because prior efforts fail to answer 
how to \textit{share} compute and storage resources of the hardware to guarantee performance 
when not all neurons and synapses of an SNN can fit on the hardware at once. 
\textcolor{black}{Table~\ref{tab:apps}, shown in Section~\ref{sec:evaluation}, lists the number of neurons and synapses in standard machine learning applications, which are on the order of thousands of neurons and hundreds of thousands of synapses.} A neuromorphic hardware such as DYNAP-SE \cite{Moradi_etal18} has four crossbars and each crossbar can accommodate a maximum of 128 fanin synapses per neuron.
Clearly, the four crossbars 
must be time-shared when executing an SNN model, which can %
lead to 
lower performance. 

\begin{figure}[h!]
	\centering
	\vspace{-5pt}
	\centerline{\includegraphics[width=0.99\columnwidth]{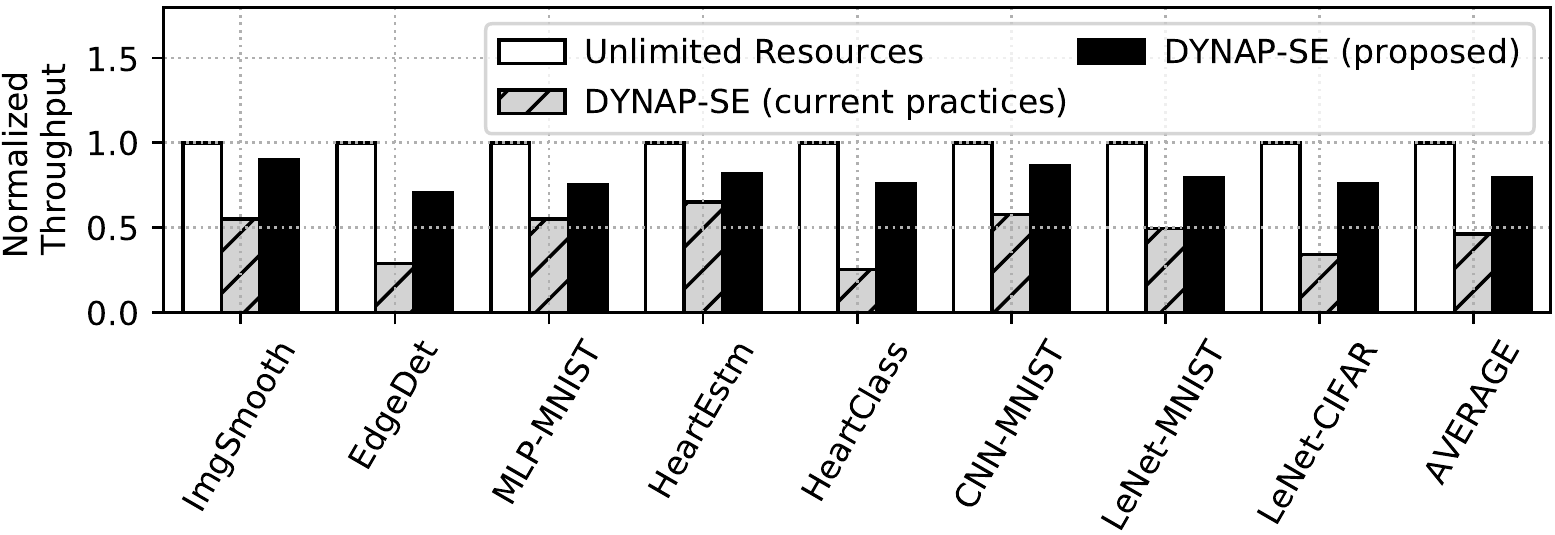}}
	\vspace{-10pt}
	\caption{Performance impact due to limited resources.}
	\vspace{-10pt}
	\label{fig:motivation}
\end{figure}

Figure~\ref{fig:motivation} illustrates the throughput impact due to limited resources on DYNAP-SE for the evaluated applications (see Section~\ref{sec:evaluation}).
We observe that throughput obtained on DYNAP-SE using current practices is on average 64\% lower than throughput analyzed using unlimited resources. 
\emph{Our objective is to reduce this performance gap when compiling SNN-based applications on neuromorphic hardware with limited resources.} 
\mc{
The figure also plots our approach, which achieves an average 78\% higher throughput than current practices.
}

A second limitation of existing approaches is that they do not address run-time aspects, i.e., how to \textit{compile and admit} machine learning applications to the hardware in the \emph{least possible time} based on the available resources.

To address these limitations, we propose a systematic and predictable approach to compile and map SNN-based machine learning applications on resource-constrained neuromorphic hardware, providing performance guarantee. 

\textbf{Contributions:} Following are our key contributions.
\begin{itemize}
    \item We propose to partition an SNN into clusters of neurons and synapses, where each cluster can fit on to the resources of a crossbar in the hardware. We pose this as a bin-packing problem and propose a greedy strategy to solve it, maximizing crossbar utilization. 
    \item We exploit the rich semantics and expressiveness of Synchronous Data Flow Graphs (SDFGs) to represent a clustered SNN and use Max-Plus Algebra to analyze its performance, e.g., throughput.
    \item We model resource constraints such as limited crossbars, input and output buffer sizes, and communication channel bandwidth into the SDFG representation. We extend the Max-Plus Algebra and use Self-Timed Execution to construct static-order schedules to estimate performance of this hardware-aware SDFG.
    \item We exploit a property of Self-Timed Scheduling to derive the schedule for each tile at run-time, starting from a single static-order schedule, without having to construct these schedules from scratch. This reduces the time to compile and admit a machine learning application to the hardware at run time and adapt dynamically to the available hardware resources.
\end{itemize}

\begin{figure}[h!]
	\centering
	\vspace{-5pt}
	\centerline{\includegraphics[width=0.99\columnwidth]{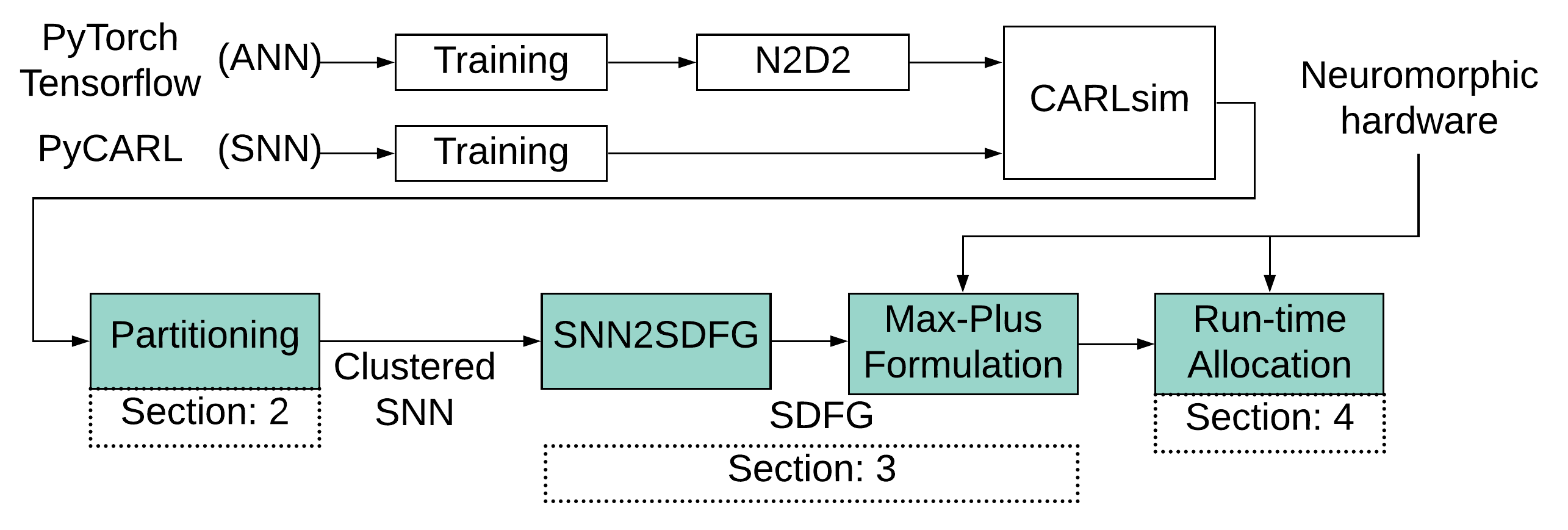}}
	\vspace{-10pt}
	\caption{Our proposed approach.}
	\vspace{-10pt}
	\label{fig:proposed_approach}
\end{figure}

Figure~\ref{fig:proposed_approach} shows a high-level workflow of our proposed approach. 
The colored boxes in this figure are our key contributions.
\mc{This workflow incorporates both Artificial Neural Network (ANN)-based applications written in a high-level language such as PyTorch \cite{paszke2019pytorch} or Tensorflow \cite{abadi2016tensorflow}, and SNN-based applications written in PyCARL \cite{balaji20pycarl}. In the former scenario, analog computations of the trained ANN-based model is first converted into spiking domain using the \texttt{N2D2} (Neural Network Design \& Deployment) tool~\cite{bichler2017n2d2}, an open-source framework for Deep Neural Network (DNN) simulation and full SNN-based applications building. Using this tool we have previously demonstrated conversion of the heart-rate classification application \cite{balaji2018power}, with less than 5\% accuracy loss. Once an SNN-based application is available, we simulate the model in CARLsim \cite{ChouCARLsim4} to record the number of spikes for each neuron in the model when excited with the training input. This spike information is then used in the partitioning step of our workflow to generate a clustered model, which is then used for analyzing performance on hardware.
}


\vspace{-5pt}

We evaluate performance and scalability of our approach using standard SNN-based applications. Our results, detailed in Section \ref{sec:results}, demonstrate a significant performance improvement compared to standard practices.

\section{Crossbar-Aware Clustering of SNNs}
\label{sec:clustering}
\subsection{Introduction to Spiking Neural Networks}
\label{sec:snn}
An SNN is a computation model with spiking neurons and synapses.  
Neurons communicate with each other by sending short impulses of infinitesimally small duration, called {spikes}, via synapses. 
Spiking neurons can be organized into feedforward topology, which consists of one input layer, one or more hidden layers, and one output layer (e.g., DNN~\cite{lecun2015deep}). Spiking neurons can also be organized in a recurrent topology~\cite{maass2002real}.
SNN-based machine learning applications, especially those that are deployed on sensor and edge devices of an IoT, typically operate on streaming data, i.e., these applications are \textbf{iterative} in nature. For these applications, real-time performance is measured in terms of \textbf{throughput}. We formulate throughput in Section~\ref{sec:perf}.

\subsection{Crossbar Resource Constraints}
A typical neuromorphic hardware (see Figure~\ref{fig:neuromorphic_hardware}) consists of crossbars, which are interconnected using an interconnection fabric. A crossbar implements neuron dynamics and facilitates synaptic storage. Therefore, each neuron and synapse of an SNN must be mapped to one of these crossbars.

\begin{figure}[h!]
	\centering
	\vspace{-10pt}
	\centerline{\includegraphics[width=0.9\columnwidth]{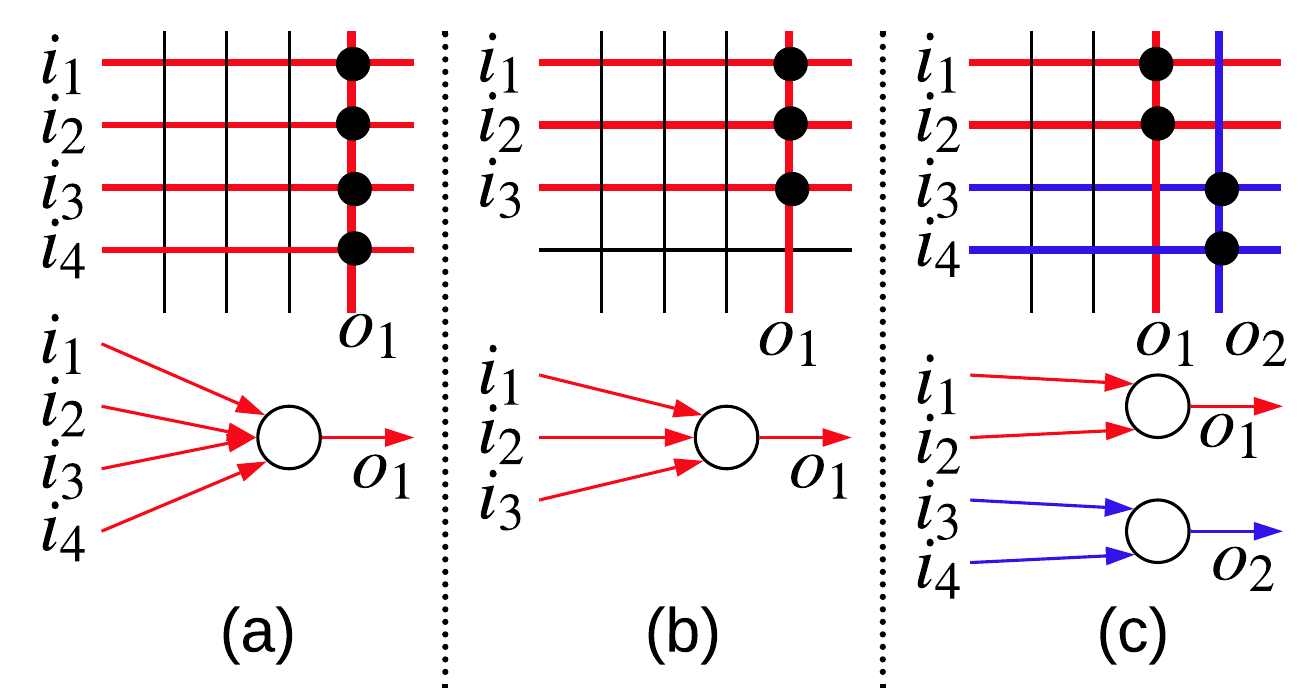}}
	\vspace{-10pt}
	\caption{Mapping of neurons \& synapses to a 4x4 crossbar.}
	\vspace{-10pt}
	\label{fig:crossbar_mapping}
\end{figure}

In terms of constraints, a crossbar can accommodate only a fixed number of synapses per neuron. This is illustrated in Figure~\ref{fig:crossbar_mapping} with three examples using a small $4 \times 4$ crossbar. In Figure~\ref{fig:crossbar_mapping}(a), the crossbar implements a single 4-input neuron. In this example, 5 out of 8 (62.5\%) input and output (IO) ports are utilized, and 4 out of 16 (25\%) crosspoints are utilized. In Figure~\ref{fig:crossbar_mapping}(b), the crossbar implements one 3-input neuron; the IO and crosspoint utilization are 50\% and 18.75\%, respectively. Finally, in Figure~\ref{fig:crossbar_mapping}(c), the crossbar implements two 2-input neurons, resulting in IO and crosspoint utilization of 75\% and 25\%, respectively. Clearly, utilization varies based on how neurons and synapses of an SNN are mapped to a crossbar. 

\begin{figure}[h!]
	\centering
	\vspace{-5pt}
	\centerline{\includegraphics[width=0.99\columnwidth]{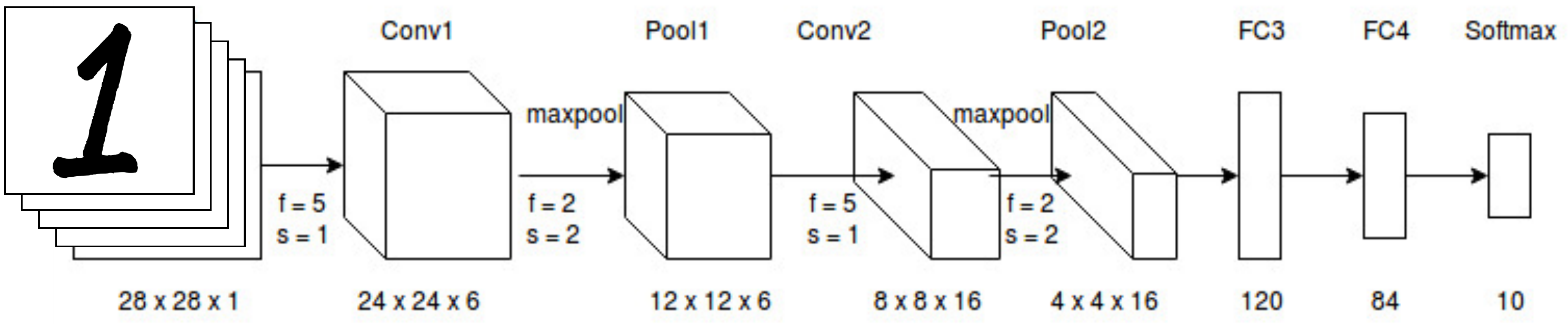}}
	\vspace{-10pt}
	\caption{The LeNet Cnvolutional Neural Network used for handwritten digit classification.}
	\vspace{-10pt}
	\label{fig:lenet}
\end{figure}

The SNN of a machine learning application can have many neurons with many synapses per neuron. Take the example of LeNet~\cite{lecun2015lenet}, a state-of-the-art convolutional neural network (CNN) to classify handwritten digits (Figure~\ref{fig:lenet}). This application has 
4,634 neurons and 1,029,286 synapses, much beyond what a single crossbar can accommodate. To map such a large SNN to the hardware, 
the SNN needs to be partitioned 
into clusters of neurons and synapses, where each cluster can fit on to the resources of a crossbar in the hardware. We discuss how to form clusters from an SNN in Sec.~\ref{sec:SNN_partitioning} and how to share crossbars among clusters in Sec.~\ref{sec:dataflow}.

\subsection{SNN Partitioning}
\label{sec:SNN_partitioning}
The SNN partitioning problem is a classic bin-packing problem and we propose a greedy strategy to solve this. Algorithm \ref{alg:clustering} shows the pseudo-code of this clustering algorithm. We first sort (in ascending order) neurons based on each neuron's fanin synapses and store them in a list (\texttt{neuron\_list}). For each neuron in this sorted list, we check to see if this neuron can be merged in one of the existing clusters in the \texttt{cluster\_list}. \mc{A neuron can be merged in a cluster if the total number of IOs, crosspoints, and buffer usage of the cluster after merging the neuron can still fit on a crossbar of the hardware. This is to ensure that the clustered SNN is deadlock-free when executed on the hardware.} If the neuron can be merged, we assign the neuron and its fanin synapses to the cluster. Otherwise, we form a new cluster. The \texttt{cluster\_list} is sorted in descending order of utilization so that the less utilized clusters can be used for merging neurons with higher fanin.

\vspace{-10pt}
\begin{algorithm}[h]
	\scriptsize{
		\texttt{neuron\_list} = sort neurons of the SNN based on their fanin synapses\;
		\texttt{clusters\_list} = \{\}\;
		\ForEach{$n\in\texttt{neuron\_list}$}{
		    find $C \in \texttt{cluster\_list}$ such that $n$ can be merged in $C$\tcc*[r]{A neuron can be merged to a cluster if the IO, bandwidth, and buffer constraints of the cluster post merging are not violated.}
		    \If{$C = \emptyset$}{
		        \texttt{cluster\_list.push($C$)}\;
		    }\Else{
		        Assign $n$ to $C$\;
		    }
		    sort \texttt{cluster\_list} in descending order of IO and crosspoint utilizations\;
		}
		check for consistency, connectivity, and deadlock in the clustered SNN\;
	}
	\caption{\small Crossbar-aware SNN partitioning.}
	\label{alg:clustering}
\end{algorithm}
\vspace{-10pt}

\subsection{Analyzing Inter-cluster Communication}
\label{sec:inter_cluster_communication}
\mc{
After partitioning, we analyze the inter-cluster communication, i.e., the number of spikes that are expected between these clusters when an SNN model is deployed in the field on a neuromorphic hardware. 
We use the spike information collected during CARLsim-based SNN simulation (Fig.~\ref{fig:proposed_approach}) to
compute the number of spikes between each cluster pair using the neuron-to-cluster mapping obtained from Algorithm~\ref{alg:clustering}. 
Next, we describe how to analyze this clustered SNN.
}

\vspace{-10pt}
\section{Dataflow Modeling of SNN Clusters}
\label{sec:dataflow}
We model a clustered SNN as a Synchronous Data Flow Graph (SDFG) for predictable performance analysis.
\subsection{Operational Semantics of SDF Graphs}
\label{sec:sdfg}
Synchronous Data Flow Graphs (SDFGs, see \cite{lee1987synchronous}) are commonly used to model streaming applications that are implemented on a multi-processor system-on-chip~\cite{SB00}. Both pipelined streaming and cyclic dependencies between tasks can be easily modeled in SDFGs. These graphs are used to analyze a system in terms of throughput and other performance properties, e.g. execution time and buffer requirements~\cite{Stuijk06dac}. 

Nodes of an SDFG are called \textit{actors}. Each node is a cluster of the SNN. Actors are computed by reading \textit{tokens} (spikes) from their input ports and writing the results of the computation as tokens on the output ports. The number of tokens produced or consumed in one execution of an actor is called the \textit{port rate}. They represent the number of spikes per unit time at the input and output of different clusters in the SNN. Port rates are visualized as annotations on edges. Actor execution is also called \textit{firing}, and it requires a fixed amount of time to execute on a crossbar. Edges in the graph are called \textit{channels} and they represent dependencies among actors.

\begin{figure}[h!]
	\centering
	\vspace{-5pt}
	\centerline{\includegraphics[width=0.7\columnwidth]{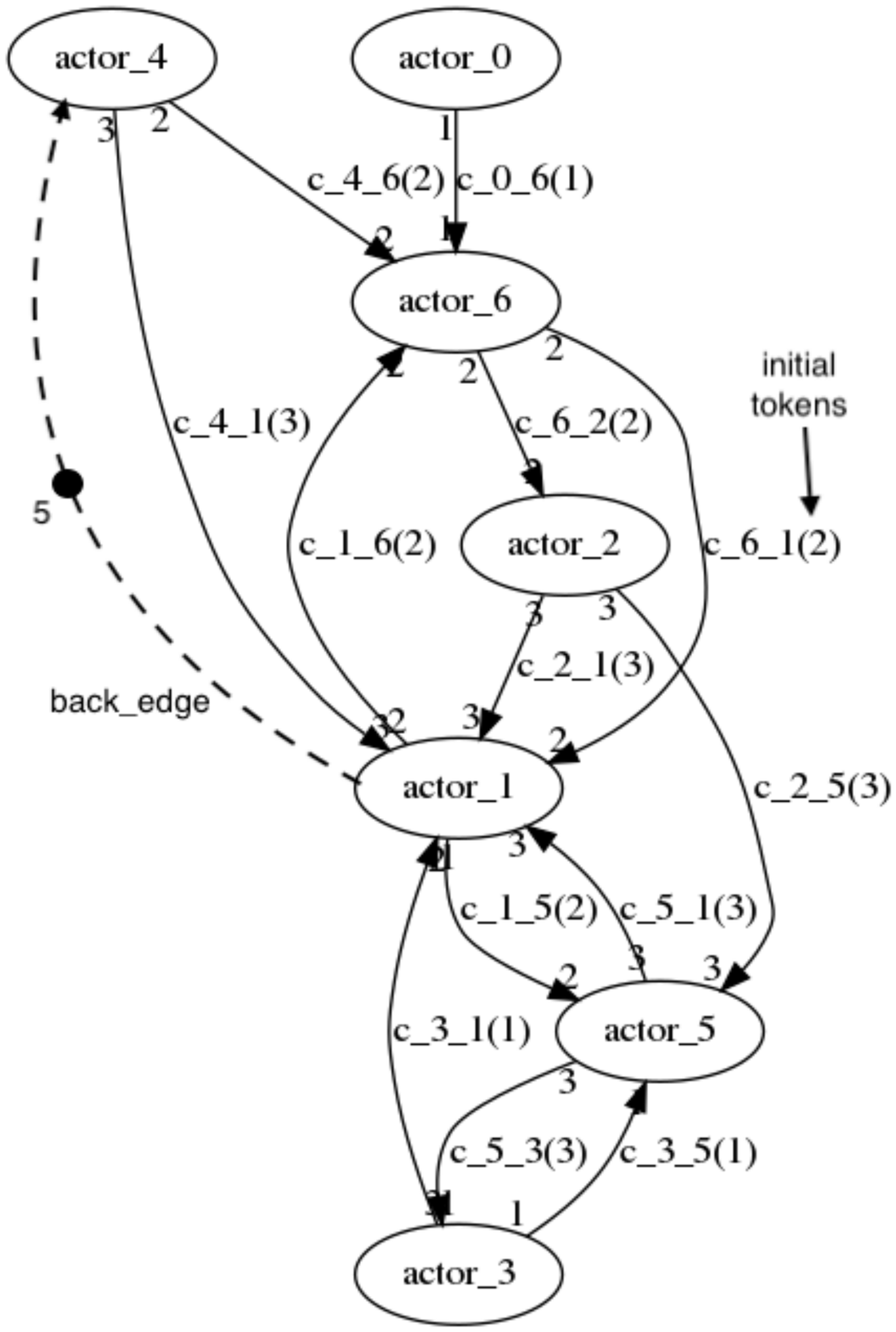}}
	\vspace{-10pt}
	\caption{SDFG representation of the LeNet CNN model used for handwritten digit classification \cite{lecun2015lenet}.}
	\vspace{-10pt}
	\label{fig:sdfg_example}
\end{figure}

\begin{figure}[h!]
	\centering
	\vspace{-5pt}
	\centerline{\includegraphics[width=0.99\columnwidth]{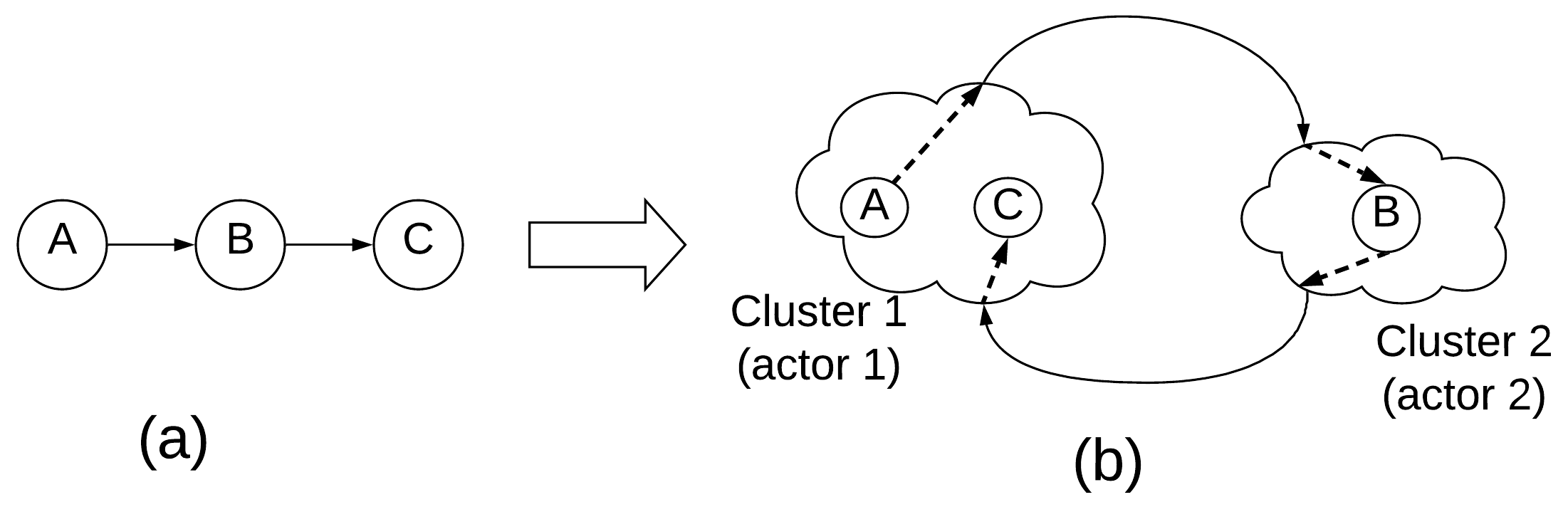}}
	\vspace{-10pt}
	\caption{An example cycle generated due to partitioning.}
	\vspace{-10pt}
	\label{fig:cycle_example}
\end{figure}

Figure~\ref{fig:sdfg_example} shows the example of an SDFG constructed using our \textcolor{purple}{SNN2SDF} tool~\cite{snn2sdf_simulator} for the LeNet CNN model used in handwritten digit classification~\cite{lecun2015lenet}.
There are 7 actors and 13 channels in this graph. For instance, \texttt{actor\_4} has two outgoing channels. The channel going to \texttt{actor\_6} has a port rate of 2 spikes per unit time, and the one going to \texttt{actor\_1} has a port rate of 3 spikes per unit time. 
From Fig.~\ref{fig:sdfg_example} we also see that there are cycles in the graph. 
\mc{
Such cycles may arise during the partitioning step. Figure \ref{fig:cycle_example}(a) illustrates a simple feedforward network of 3 neurons (A, B, \& C). Figure \ref{fig:cycle_example}(b) illustrates a scenario where neurons A and C are placed in cluster 1 (actor 1) and neuron B in cluster 2 (actor 2) during partitioning. Due to the connectivity of the neurons in Figure \ref{fig:cycle_example}(a), there is a cyclic dependencency between the two actors: \underline{\texttt{actor\_1}$\rightarrow$\texttt{actor\_2}$\rightarrow$\texttt{actor\_1}}.
Therefore, Directed Acyclic Graphs (DAGs) \textit{cannot} be used to represent and analyze clustered SNNs. This justifies our choice of using SDFGs. 
}

An actor is called {\em ready} when it has sufficient input tokens on all its input channels and sufficient buffer space on all its output channels; an actor can only fire when it is ready. A channel may also contain an \textit{initial~token}, shown as annotation. For instance, the channel between \texttt{actor\_0} and \texttt{actor\_6} in the figure has 1 initial token.
A set $Ports$ of ports is assumed, and with each port $p \in Ports$, a finite rate $Rate(p) \in \mathbb{N}\setminus\{0\}$ is associated. Formally, 

\vspace{-5pt}

\begin{Definition}{Actor}
{An actor $\actor{a}_i$ is a tuple $(I_i,O_i,\tau_i,\mu_i)$ consisting of a set $I_i$ ($\subseteq Ports$) of input ports, a set $O_i$ ($\subseteq Ports$) of output ports with $I_i \cap O_i = \emptyset$, $\tau_i$ is the execution time of $\actor{a}_i$ and $\mu_i$ is its state space, i.e., buffer space needed for communicating spikes on all of its channels.}
\end{Definition}

\vspace{-10pt}

\begin{Definition}{SDFG}
{An SDFG is a directed graph $G_{app} = (A,C)$ consisting of a finite set $A$ of actors and a finite set $C\subseteq Ports^2$ of channels. The source of channel $ch_i^j \in C$ is an output port of actor $\actor{a}_i$, the destination is an input port of actor $\actor{a}_j$. All ports of all actors are connected to precisely one channel, and all channels are connected to ports of some actors. The source and the destination port of channel $ch_i^j$ are denoted by $SrcP(ch_i^j)$ and $DstP(ch_i^j)$ respectively. Channels connected to the input and output ports of an actor $\actor{a}_i$ are denoted by $InC(\actor{a}_i)$ and $OutC(\actor{a}_i$) respectively.}
\end{Definition}

Before an actor $\actor{a}_i$ starts its firing, it requires $Rate(q_i)$ tokens from all $(p,q_i)\in InC(\actor{a}_i)$. When the actor completes execution, it produces $Rate(p_i)$ tokens on every $(p_i,q) \in OutC(\actor{a}_i)$. One important property of an SDFG is \textit{throughput}, which is defined as the inverse of its long-term period. A period is the average time needed for one iteration of the SDFG. An iteration is defined as the minimum non-zero execution such that the original state of the SDFG is obtained. This is the performance parameter used in this paper. Following definitions are introduced to formulate throughput.

\begin{Definition}{Repetition Vector}
The Repetition Vector \emph{RptV} of an SDFG is defined as the vector specifying the number of times actors in the SDFG are executed in one iteration.
\end{Definition}
In the SDFG representation of a clustered SNN, 
all spikes generated on a channel are consumed by the destination actor. This means that all actors are fired exactly once during one iteration of the application. So, $RptV = [1 1 1 1 1 1 1]$. 
\mc{
Further\-more, by incorporating constraints of a crossbar during the partitioning, we ensure that the SDFG generated from the clustered SNN is consistent, connected, and deadlock free.
}

\subsection{Computing Performance on Infinite Resources}
\label{sec:perf}
We present an approach to compute the application period of an SDFG by analyzing its maximum cycle mean (MCM) and assuming infinite hardware resources. 
\mc{
For this, we use Max-Plus Algebra \cite{heidergott2014max,zhang2013sdc,cong2006efficient}. 
The key difference of our approach with these prior approaches is the incorporation of resource constraints, which we describe next.
}
The Max-Plus semiring $\mathbb{R}_{\text{max}}$ is the set $\mathbb{R}\cup\{-\infty\}$ defined with two basic operations $\oplus \text{ and } \otimes$, which are related to linear algebra as
\begin{equation}
\label{eq:mpb}
\footnotesize a \oplus b = \max(a,b) \text{  and  } a \otimes b = a + b
\end{equation}

To use Max-Plus Algebra to analyze an SDFG, it is customary to express the time at which an actor fires in terms of preceding firings and then use standard analysis techniques for Max-Plus Algebra to estimate timing performance. For the SDFG in Figure~\ref{fig:sdfg_example}, firing end time of all 7 actors in the $k^{\text{th}}$ iteration (in linear algebra) are 

\vspace{-10pt}
\begin{footnotesize}
\begin{align}
\label{eq:laeqn}
t_{0}(k) &\ge t_{0}(k-1) + \tau_{0}\\
t_{1}(k) &= \max\Big[t_3(k-1),t_5(k-1),t_{2}(k),t_{4}(k),t_{6}(k)\Big] + \tau_{1}\nonumber\\
t_{2}(k) &= t_{6}(k) + \tau_{2}\nonumber\\
t_{3}(k) &= t_5(k) + \tau_{3}\nonumber\\
t_{4}(k) &\ge t_{4}(k-1) + \tau_{4}\nonumber\\
t_{5}(k) &= \max\Big[t_3(k-1),t_{1}(k),t_2(k)\Big] + \tau_{5}\nonumber\\
t_{6}(k) &= \max\Big[t_1(k-1),t_{0}(k),t_4(k)\Big] + \tau_{6}\nonumber
\end{align}
\end{footnotesize}\normalsize

Observe that the firing end time of actor \texttt{actor\_i} in the  $k^\text{th}$ iteration is after its firing end time in the $(k-1)^\text{th}$ iteration. Furthermore, the production and consumption rates are the same for every channel in the SDFG. Using previously introduced Max-Plus semantics, firing end times for every actor in the SDFG  can be expressed as
\begin{equation}
\label{eq:mpaeq}
\footnotesize t_\mathfrak{n}(k) = \oplus t_\mathfrak{m}(k-1) \otimes \tau_\mathfrak{n}\text{, }\forall \mathfrak{m}\in Pre(\mathfrak{n})
\end{equation}
With a simple transformation of variables, the above sum-of-product equation can be rewritten as 
\begin{equation}
\label{eq:mat}
\footnotesize\mathbf{t_k} = \mathbf{{T}\cdot t_{k-1}}
\end{equation}
where $\mathbf{{T}}$ captures execution times $\tau_\mathfrak{n}$. The following definitions are introduced to estimate latency.

\vspace{-5pt}

\begin{Definition}{{Digraph}}
	The digraph $\Gamma(T)$ of a $n\times n$ matrix $T$ with entries defined in $\mathbb{R}_{\text{max}}$ is the tuple $\langle A,E\rangle$, where $A$ is the set of vertices, i.e., $A = \{1,2,\cdots n\}$ and $E$ is the set of connected ordered arcs between vertices i.e., $E = \{(i,j)~|~T_{i,j}\neq -\infty\}$.
\end{Definition}

\vspace{-5pt}

\begin{Definition}{{Walk}}
	A walk $w$ in digraph $\Gamma(T)$ is the sequence of arcs $(x_1,x_2)(x_2,x_3)\cdots(x_{k-1},x_k)$;  head of an arc in the sequence is either the start vertex of the walk or tail vertex of a preceding arc; and the tail vertex of an arc in the sequence is either the end vertex of the walk or head vertex of a succeeding arc. Weight of the walk is given by
	\begin{equation}
	\label{eq:weight}
	\footnotesize|w|_T =  T_{x_1 x_2} + \cdots T_{x_{k-1} x_k}
	\end{equation}
\end{Definition}

\begin{Definition}{{Cycle}}
	A cycle $c$ in digraph $\Gamma(T)$ is the walk $(x_1,x_2)(x_2,x_3)\cdots(x_{k-1},x_k)$, such that $x_k = x_1$.
\end{Definition}

\begin{Definition}{{Maximum Cycle Mean}}
	The maximum cycle mean, $\rho_\text{max} (T)$ is the maximum of the weight-to-length ratio of all cycles $c$ in $\Gamma(T)$ i.e.,
	\begin{equation}
	\label{eq:mcm}
	\footnotesize\rho_\text{max} (T) = \max\limits_{\forall c \text{ in }\Gamma(T)}\frac{|c|_T}{|c|} = \max\limits_{k\geq 1} \max\limits_{x_1,\cdots,x_k} \frac{T_{x_1 x_2} + \cdots T_{x_k x_1}}{k}
	\end{equation}
\end{Definition}

In this paper, \textbf{performance of an SNN is defined in terms of throughput} of the equivalent SDFG, measured as the inverse of its \textit{maximum cycle mean} (Equation~\ref{eq:mcm}).

\section{Hardware-Aware Performance Analysis}
\label{sec:performance}
We now 
extend the Max-Plus formulation to analyze performance of an SNN on a resource-constrained hardware.
\subsection{Platform Description}
\begin{figure}[h!]
	\centering
	\vspace{-5pt}
	\centerline{\includegraphics[width=0.9\columnwidth]{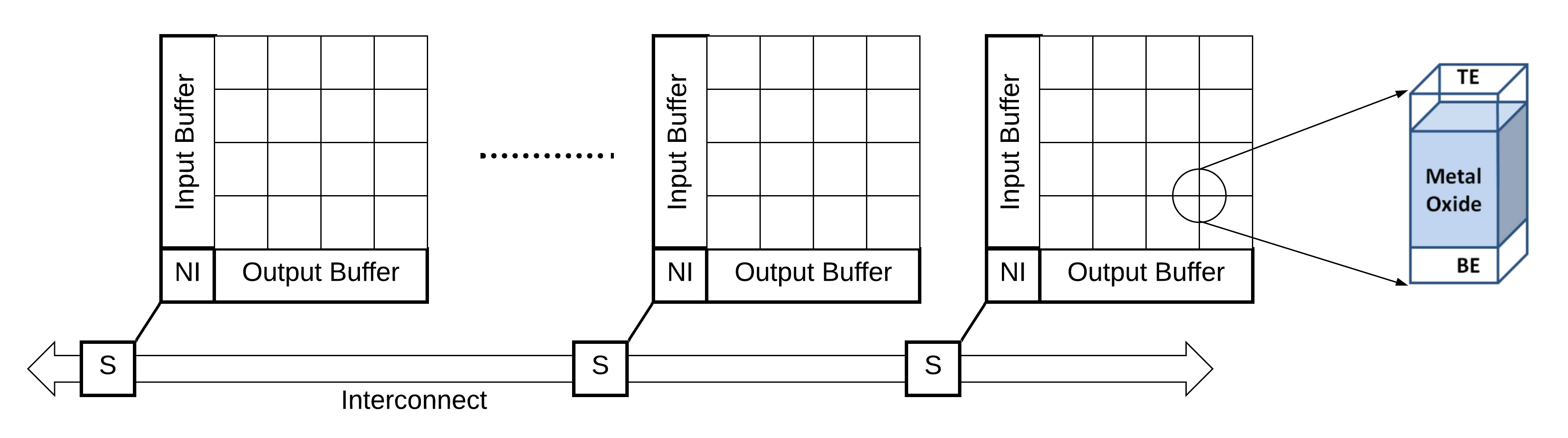}}
	\vspace{-10pt}
	\caption{An example tile-based neuromorphic hardware. Each tile consists of a crossbar, input and output buffers, and a network interface (NI).} 
	\vspace{-10pt}
	\label{fig:neuromorphic_hardware}
\end{figure}

Performance of an SNN, computed using Equation~\ref{eq:mcm}, gives the maximum period possible with infinite hardware resources in terms of crossbars, buffer sizes, and communication bandwidth. For off-the-shelf neuromorphic hardware, however, these resources are limited. Figure~\ref{fig:neuromorphic_hardware} shows a typical tile-based neuromorphic hardware, where tiles are connected via an interconnection. 
Each tile consists of a crossbar (C), input and output buffers, and a network interface (NI). A crossbar is a two dimensional organization of horizontal and vertical electrodes. At every cross-point, there is a 
Oxide-based Resistive RAM (OxRAM)~\cite{garbin2015hfo} for synaptic storage.

\subsection{Binding Actors (Clusters) to Tiles}
\label{sec:binding}
Similar in vein to PYCARL~\cite{balaji20pycarl}, we use a load balancing strategy to bind clusters of an SNN to the tiles of the hardware. We first formulate the load of a tile as follows:
\begin{equation}
    \label{eq:tile_load}
    \footnotesize load(tile) = a*\text{crossbar} + b*\text{buffer} + c* \text{connection} + d*\text{bandwidth}
\end{equation}
where \ineq{a,b,c} and \ineq{d} are user-defined constants used to prioritize different hardware resources on a tile.
Next, we propose a greedy approach to balance the load on each tile. For this, we first distribute the clusters evenly to the tiles and calculate the standard deviation of tile loads. For every cluster pair that is bound to two different tiles, we swap the clusters to see if the standard deviation reduces. If it reduces, we retain this new binding and continue analyzing other cluster pairs. 

\subsection{Executing a Cluster on a Crossbar}
\begin{figure}[h!]
	\centering
	\vspace{-5pt}
	\centerline{\includegraphics[width=0.8\columnwidth]{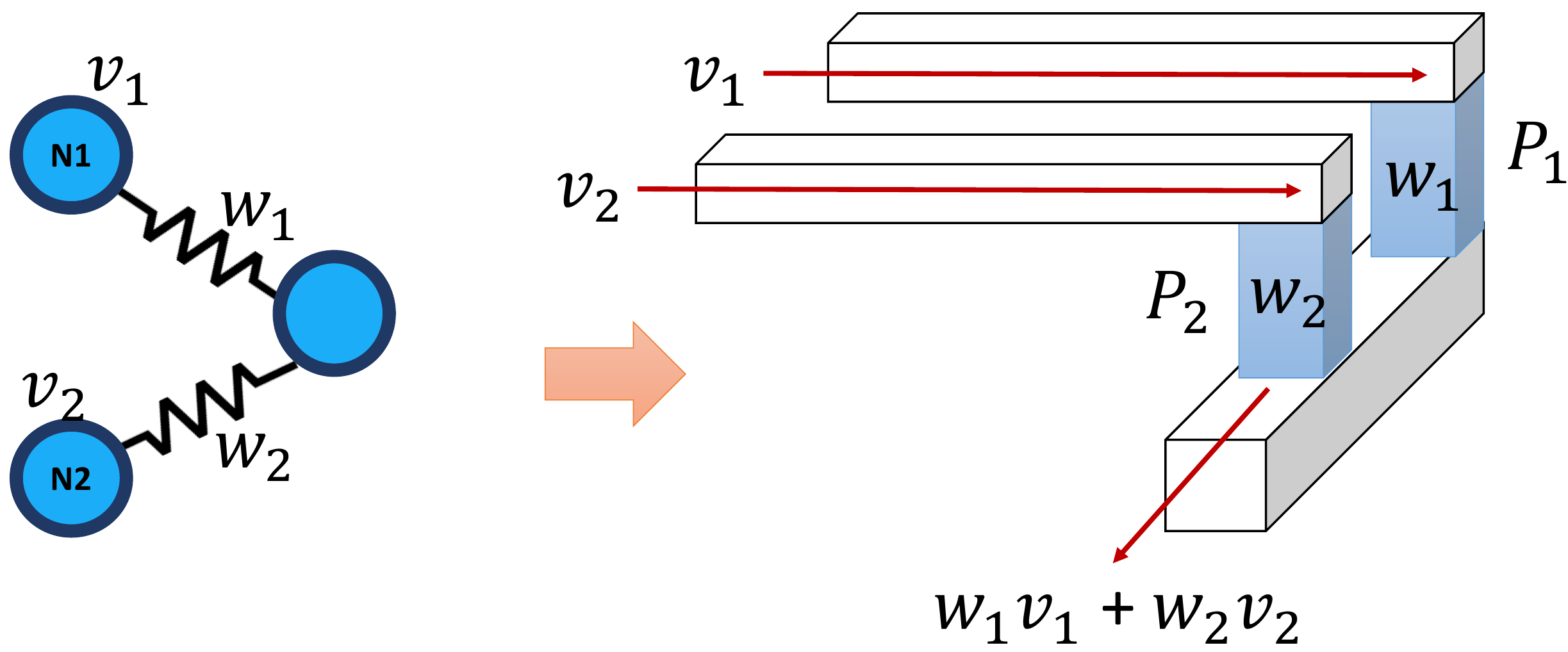}}
	\vspace{-10pt}
	\caption{Executing a cluster on a crossbar.}
	\vspace{-10pt}
	\label{fig:oxram_mapping}
\end{figure}

A cluster is executed by placing its neurons and synapses on to the crossbar of a tile. Figure~\ref{fig:oxram_mapping} illustrates this execution mechanism. Synaptic weights \ineq{w_{1}} and \ineq{w_{2}} are programmed into OxRAM cells P1 and  P2, respectively. The output spike voltages, \ineq{v_1} from N1 and \ineq{v_2} from N2, inject current into the crossbar, which is obtained by multiplying a pre-synaptic neuron's output spike voltage with the OxRAM cell's conductance at the cross-point of the pre- and post-synaptic neurons (following Ohm's law). Current summations along columns are performed in parallel using Kirchhoff’s current law, and implement the sums $\sum_j w_{i}v_i$, needed for forward propagation of neuron excitation. 
The execution time of a cluster is the current propagation delay through an OxRAM synapse and is obtained from Malik et al.~\cite{mallik2017design}.
Note that though we use OxRAM synapses as example, 
the execution technique
applies to any resistive non-volatile synapse.

\begin{figure}[h]
	\centering
	\vspace{-5pt}
	\centerline{\includegraphics[width=0.8\columnwidth]{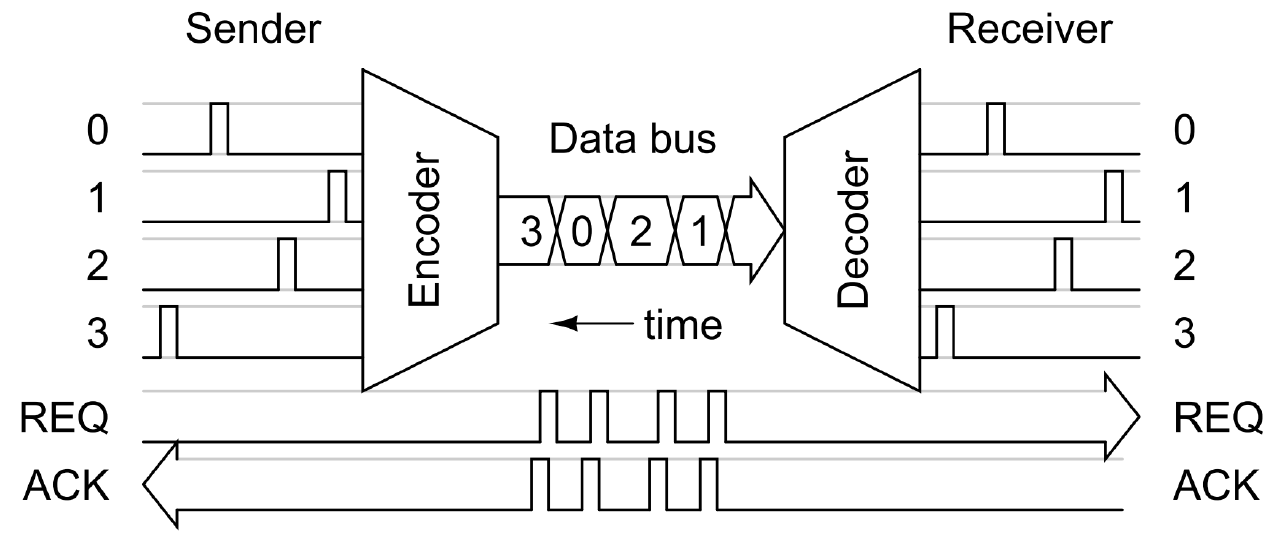}}
	\vspace{-10pt}
	\caption{An example AER protocol (adapted from~\cite{boahen1998communicating}).}
	\vspace{-10pt}
	\label{fig:aer}
\end{figure}

Although a crossbar implements analog computations, spikes at the output are converted into digital packets before communicating on the interconnect. We use the Address Event Representation (AER) protocol~\cite{boahen1998communicating}.
Figure~\ref{fig:aer} shows an example explaining the principles behind AER. Here, four neurons in a crossbar spikes at time 3, 0, 1 and 2 time units, respectively. The encoder encodes these four spikes in order to be communicated on the interconnect.  As can be clearly seen from this figure, a spike is encoded uniquely with its source and time of spike. Therefore, each token in the SDFG is simply a spike packet with header encoding the address and time, and zero payload.

 \subsection{Computing-Resource Aware Performance}
{To compute performance of an SNN on a resource-constrained neuromorphic hardware, we first construct its hardware-aware SDFG and then compute the maximum cycle mean using the Max-Plus Algebra formulation of Equation~\ref{eq:mcm}.} 

The following three steps describes
this approach
starting from the actor binding of Section~\ref{sec:binding}. Without loss of generality, we use Equation~\ref{eq:actor_allocation} as a running binding example.

\vspace{-10pt}
\begin{footnotesize}
\begin{align}
\label{eq:actor_allocation}
\texttt{\textbf{tile\_0}}: \texttt{actor\_3},\texttt{actor\_6} &~~~~~~~~\texttt{\textbf{tile\_2}}: \texttt{actor\_2}\\
\texttt{\textbf{tile\_1}}: \texttt{actor\_2},\texttt{actor\_5} &~~~~~~~~\texttt{\textbf{tile\_3}}: \texttt{actor\_1},\texttt{actor\_4} \nonumber
\end{align}
\end{footnotesize}\normalsize

\vspace{-5pt}

\textbf{Step 1 (Buffer Modeling):} Limited input and output buffer-sizes are modeled as back-edges with initial tokens in the hardware-aware SDFG. The number of tokens on this back-edge indicates the buffer-size available. When an actor generates spikes on a channel, the available size reduces; when the receiving actor consumes the spike, the available buffer is released. Figure~\ref{fig:sdfg_example} shows such an example of a back-edge, where the buffer size of the channel from \texttt{actor\_4} to \texttt{actor\_1} is shown as five. Before \texttt{actor\_4} can be executed, it has to check if enough buffer space is available. This is modeled by requiring tokens from the back-edge to be
consumed. Since it produces three tokens per firing, three tokens from the back-edge are consumed, indicating reservation of three buffer spaces. On the consumption side, when \texttt{actor\_1} is executed, it frees three buffer spaces, indicated by a release of three tokens on the back-edge. 
\mc{We assume \emph{atomic} execution of actors on a crossbar, i.e., a crossbar reads input tokens and produces output tokens in the output buffer for no more than one actor at any given instance of time. To prevent other actors mapped to the same crossbar from being fired simultaneously, 
}
the output buffer space is claimed at the start of execution and released only at the end of firing.

\begin{figure}[h!]
	\centering
	\vspace{-5pt}
	\centerline{\includegraphics[width=0.75\columnwidth]{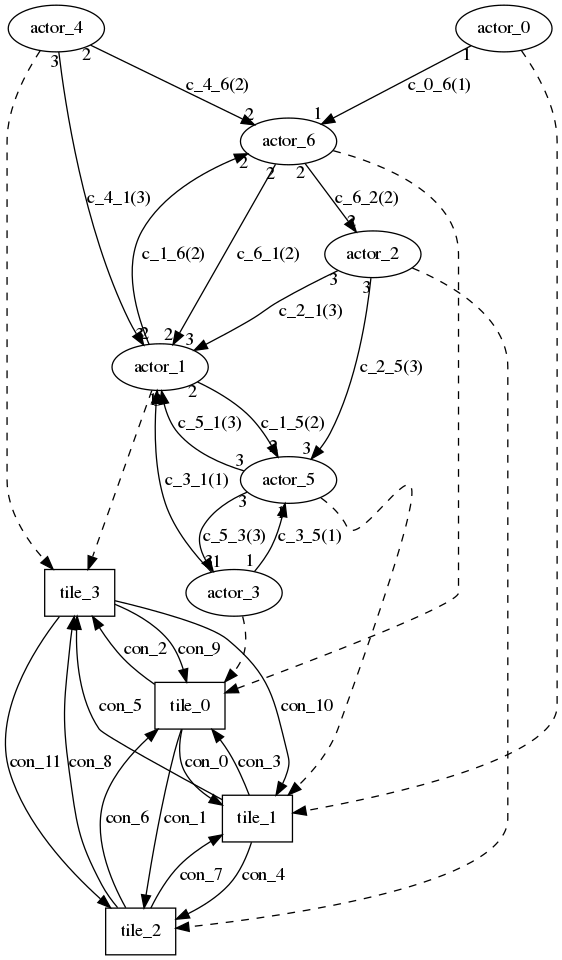}}
	\vspace{-10pt}
	\caption{Hardware-aware SDFG of LeNet-based handwritten digit recognition on the neuromorphic hardware of Fig.~\ref{fig:neuromorphic_hardware}.}
	\vspace{-10pt}
	\label{fig:platform_mapping}
\end{figure}

Figure~\ref{fig:platform_mapping} shows the final hardware-aware SDFG of LeNet-based handwritten digit classification on a neuromorphic hardware with four tiles. For simplicity of representation, we have omitted the back-edges from the figure.

\textbf{Step 2 (Actor Ordering):} 
The number of crossbars in a neuromorphic hardware is limited and therefore they may have to be shared between actors of an SNN. However, on a tile, only one instance of an actor can be executing at the same moment in time. We use time-division multiple-access (TDMA) to allocate time slices to actors mapped to the same tile. During the allocated time slice, an actor is executed on the crossbar of the tile and generates spikes, which are stored in the output buffer for communication on the interconnect. Next, we generate the order in which the actors bound to a tile are fired to provide a guarantee on performance, i.e., throughput. For this, we apply our Max-Plus Algebra formulation (Eq.~\ref{eq:mcm}) on the hardware-aware SDFG of Fig.~\ref{fig:platform_mapping}. This is our \emph{static-order schedule}. We construct this schedule at \textit{design time}.

\textbf{Step 3 (Actor Execution):} Once the static-order schedule is constructed for all tiles of the hardware, we use self-timed execution strategy~\cite{moreira2007self} for executing these actors at run-time. In this strategy, the exact firing times of actors are discarded, retaining only the assignment and ordering of actors on each tile as obtained from the design-time analysis (step 2). At run time, ready actors are inserted in a list and fired in the same order as determined from design time. 




\section{Run-time Resource Management}
\label{sec:run_time}
A modern neuromorphic hardware is expected to execute many SNN applications simultaneously. When a new application is to be admitted to a hardware, which is currently running other applications, the incoming application needs to be compiled and mapped to the hardware within a short time window, based on resources currently available on the hardware. Furthermore, when an existing application finishes execution, its hardware resources are freed, meaning that such resources can now be allocated to other running applications to improve their performance. Clearly, a dynamic compilation strategy is needed to address them. 

We observe that over 75\% of the total compilation time of an SNN application is due to the time consumed in constructing the static-order schedule for each tile of the neuromorphic hardware (see Section~\ref{sec:compile_time}). To address this, we exploit the basic property of Max-Plus Algebra and self-timed scheduling, which is expressed as the following lemma.

\vspace{-5pt}
\begin{Lemma}
	I If the schedule of actors on a single-tile system is used to derive the schedule for a multi-tile system by keeping the actor firing order unchanged, the resultant multi-tile schedule is free of deadlocks~\cite{blazewicz1976scheduling}.
\end{Lemma}
\vspace{-5pt}

\begin{figure}[h!]
	\centering
	\vspace{-5pt}
	\centerline{\includegraphics[width=0.99\columnwidth]{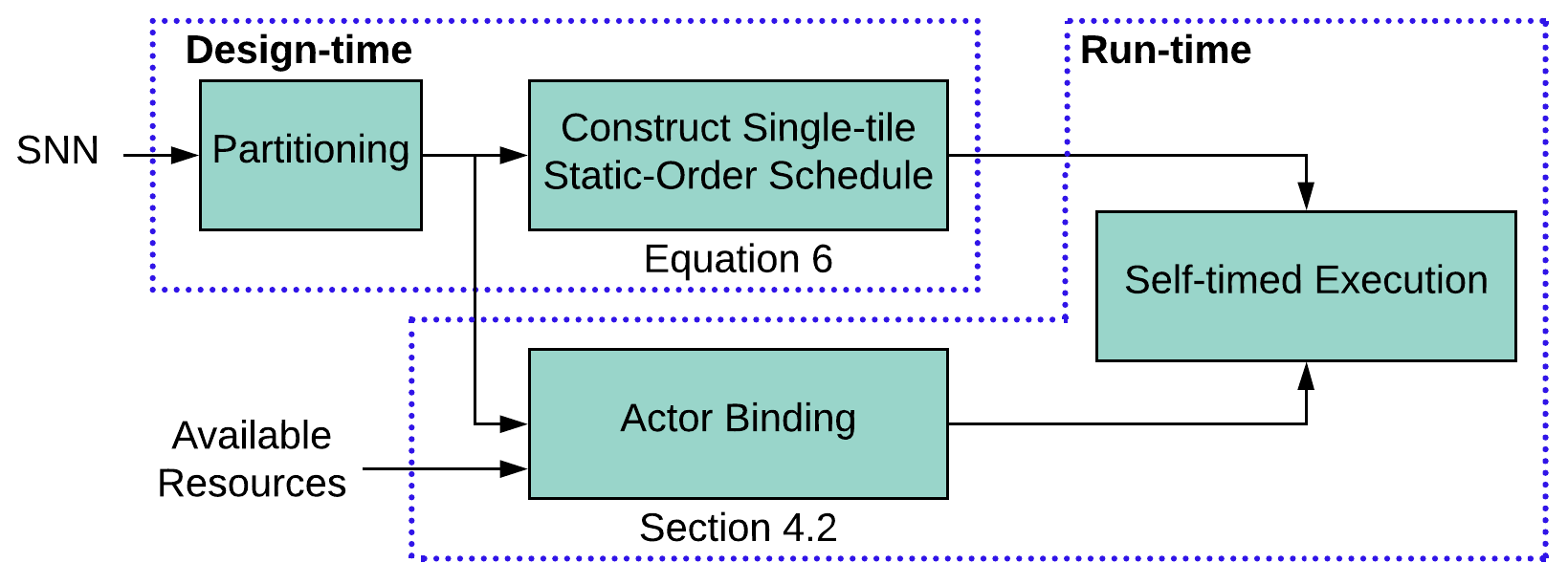}}
	\vspace{-10pt}
	\caption{Our approach to run-time resource management.}
	\vspace{-10pt}
	\label{fig:run_time}
\end{figure}

Based on this lemma, we propose the following. First, we construct the static-order schedule for all actors of an SNN on a single tile at design-time. This is achieved using our proposed Max-Plus Algebra formulation of Equation~\ref{eq:mcm}. Next, we discard the exact timing information, retaining only the actor firing orders for run-time use.
At run-time, we first construct the actor binding to tiles (Section~\ref{sec:binding}), considering the available resources. Next, we use the single-tile static-order schedule to fire actors when they are ready. Figure~\ref{fig:run_time} illustrates our run-time methodology.

\begin{figure}[h!]
	\centering
	\vspace{-5pt}
	\centerline{\includegraphics[width=0.85\columnwidth]{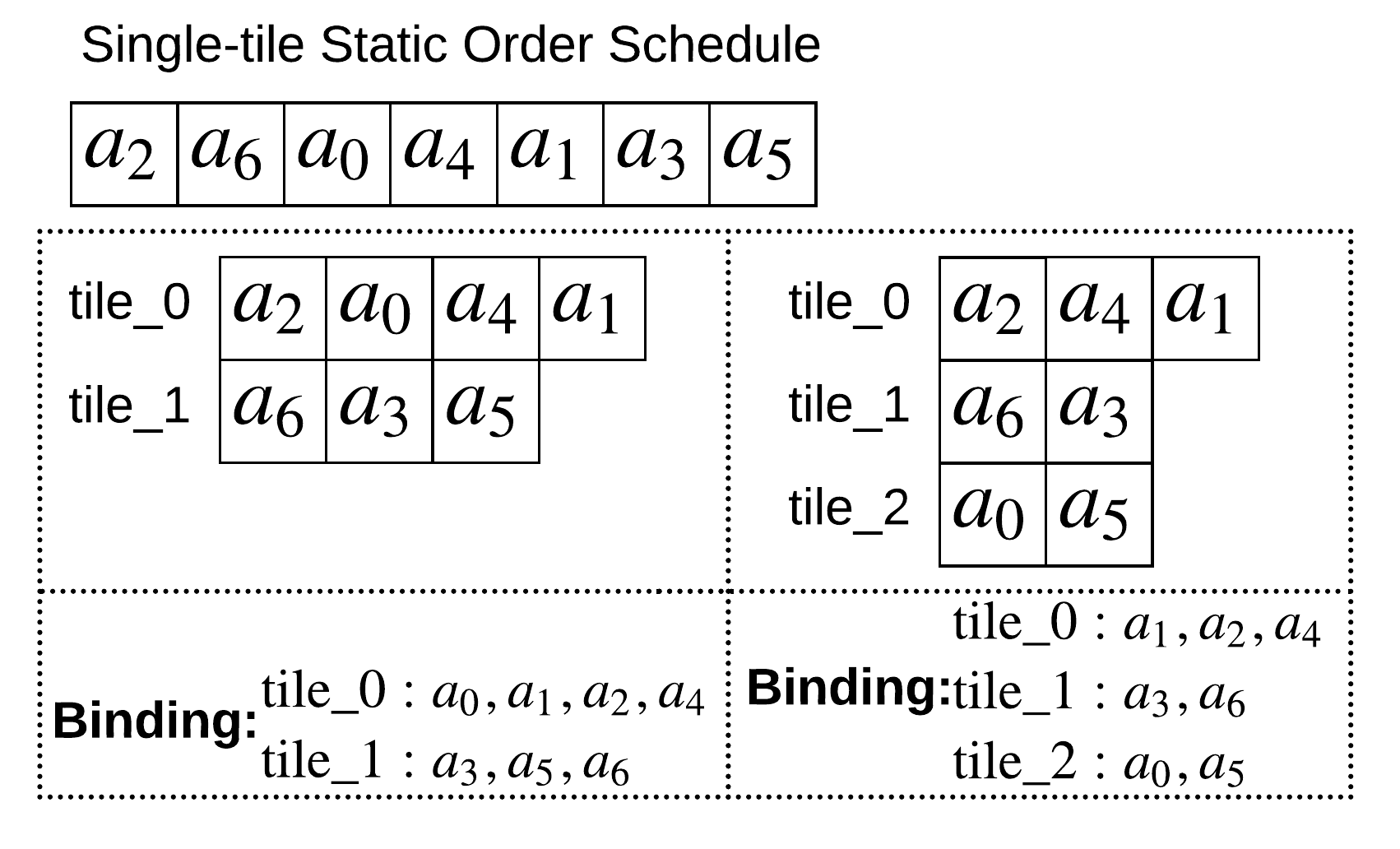}}
	\vspace{-10pt}
	\caption{Schedules constructed from the same single-tile static order schedule using 2 and 3 tiles, respectively.}
	\vspace{-10pt}
	\label{fig:schedule}
\end{figure}

Figure~\ref{fig:schedule} illustrates the construction of per-tile schedules for an SNN application with seven run-time actors, and with two different binding of actors to tiles and the same single-tile static order schedule. We illustrate two scenarios in this example. In the first scenario (left), the application uses two tiles of the hardware. In the second scenario (right), the application uses three tiles of the hardware. In both scenarios, actor orders on each tile is the same as that on the single-tile. Since tile schedules are not constructed from scratch, the schedule construction time is much lower (see Table~\ref{tab:runtime}).

However, performance obtained using this single-tile schedule can be lower than the maximum performance of a multi-tile schedule constructed independently. As long as this performance deviation is bounded, the actor schedule for any tile can be easily derived from the binding of actors to this tile and a given single-tile static-order schedule. In Section~\ref{sec:run_time}, we evaluate the performance of this scheduling.

\section{Evaluation Methodology}
\label{sec:evaluation}

\subsection{Hardware Models}
We model the DYNAP-SE neuromorphic hardware~\cite{Moradi_etal18} with the following configurations.

\begin{itemize}
    \item A tiled array of 4 crossbars, each with 128 input and 128 output neurons. There are 65,536 crosspoints (i.e., OxRAM NVMs) in each crossbar.
    \item Spikes are digitized and communicated between cores through a mesh routing network using the Address Event Representation (AER) protocol.
    \item Each synaptic element is an HfO2-Based OxRAM Device. 
    Timing parameters are modeled from~\cite{garbin2015hfo}.
\end{itemize}

To test scalability of our compilation technique, we also evaluate hardware models with 9 and 16 neuromorphic cores.

\subsection{Evaluated Applications}
We evaluate eight standard SNN-based machine learning applications: 1) {image smoothing} (ImgSmooth)~\cite{ChouCARLsim4} on $64 \times 64$ images; 2) {edge detection} (EdgeDet)~\cite{ChouCARLsim4} on $64 \times 64$ images using difference-of-Gaussian; 3) {multi-layer perceptron (MLP)-based handwritten digit recognition} (MLP-MNIST)~\cite{diehl2015unsupervised} on $28 \times 28$ images of handwritten digits from the MNIST dataset \cite{deng2012mnist}; 4) {heart-rate estimation} (HeartEstm) using electrocardiogram (ECG) data~\cite{das2018unsupervised} from the Physionet database~\cite{moody2001physionet}; 5) {ECG-based heart-beat classification} (HeartClass)~\cite{balaji2018power}; 6) {handwritten digit classification with standard CNN} (CNN-MNIST)~\cite{springenberg2014striving,reddi2019mlperf}; 7) {handwritten digit classification with the LeNet CNN} (LeNet-MNIST)~\cite{reddi2019mlperf}; and 8) {image classification with LeNet CNN} (LeNet-CIFAR)~\cite{reddi2019mlperf} with images from the CIFAR dataset~\cite{krizhevsky2014cifar}. The LeNet CNN model is described in~\cite{lecun2015lenet}.
Table~\ref{tab:apps} summarizes the topology and the number of neurons, synapses, and spikes of these applications. Image-based applications are iteratively executed on test images.

\vspace{-5pt}
\begin{table}[h!]
	\renewcommand{\arraystretch}{0.8}
	\setlength{\tabcolsep}{1.2pt}
	\centering
	\begin{threeparttable}
	{\fontsize{6}{10}\selectfont
		\begin{tabular}{l|ccl|c|c}
			\hline
			\textbf{Applications} & \textbf{Synapses} & \textbf{Neurons} & \textbf{Topology} & \textbf{Spikes} & \textbf{Accuracy}\\
			\hline
			ImgSmooth \cite{ChouCARLsim4} & 136,314 & 980 & FeedForward (4096, 1024) & 17,600 & 100\%\\
			EdgeDet \cite{ChouCARLsim4} & 272,628 &  1,372 & FeedForward (4096, 1024, 1024, 1024) & 22,780 & 100\%\\
			MLP-MNIST \cite{diehl2015unsupervised} & 79,400 & 984 & FeedForward (784, 100, 10) & 2,395,300 & 95.5\%\\
			HeartEstm \cite{das2018unsupervised} & 636,578 & 6,952 & Recurrent & 3,002,223 & 99.2\%\\
			HeartClass \cite{balaji2018power} & 2,396,521 & 24,732 & CNN\tnote{1} & 1,036,485 & 85.12\%\\
			CNN-MNIST \cite{reddi2019mlperf} & 159,553 & 5,576 & CNN\tnote{2} & 97,585 & 96.7\%\\
			LeNet-MNIST \cite{reddi2019mlperf} & 1,029,286 & 4,634 & CNN\tnote{3} & 165,997 & 99.1\%\\
			LeNet-CIFAR \cite{reddi2019mlperf} & 2,136,560 & 18,472 & CNN\tnote{4} & 589,953 & 84.0\%\\
			\hline
	\end{tabular}}
	\begin{tablenotes}\scriptsize
        \item[1.] Input(82x82) - [Conv, Pool]*16 - [Conv, Pool]*16 - FC*256 - FC*6
        \item[2.] Input(24x24) - [Conv, Pool]*16 - FC*150 - FC*10
        \item[3.] Input(32x32) - [Conv, Pool]*6 - [Conv, Pool]*16 - Conv*120 - FC*84 - FC*10
        \item[4.] Input(32x32x3) - [Conv, Pool]*6 - [Conv, Pool]*6 - FC*84 - FC*10
    \end{tablenotes}
	\end{threeparttable}
	\caption{Applications used to evaluate our approach.}
	\label{tab:apps}
\end{table}
\vspace{-24pt}

\subsection{Evaluated State-of-the-art Techniques}
We evaluate the following three approaches.
\begin{itemize}
    \item SpiNeMap~\cite{balaji2019mapping} maps SNNs to tiles, minimizing spikes on the interconnect. 
    Clusters on a tile are executed in a random order.
    \item PYCARL~\cite{balaji20pycarl} maps SNNs to tiles, balancing tile load. Clusters on a tile are executed in a random order.
    \item Our proposed approach uses SDFGs to analyze the performance of an SNN on a neuromorphic hardware. Clusters are allocated to tiles based on this analysis. Overall, our approach balances load on each tile and uses static-order schedule to improve throughput.
\end{itemize}

\subsection{Evaluated Metrics}
We evaluate the following performance metrics.


\begin{itemize}
    \item Performance: This is the throughput of each application on the hardware.
    \item Compilation Time: This is the time to compile and map each application on the hardware.
    \item Resource utilization: \mc{This is the tile, buffer, connection, and input and output bandwidth utilization on the hardware for each application.}
\end{itemize}

\section{Results and Discussion}
\label{sec:results}
\subsection{Performance}
\begin{figure}[h!]
	\centering
	\vspace{-5pt}
	\centerline{\includegraphics[width=0.99\columnwidth]{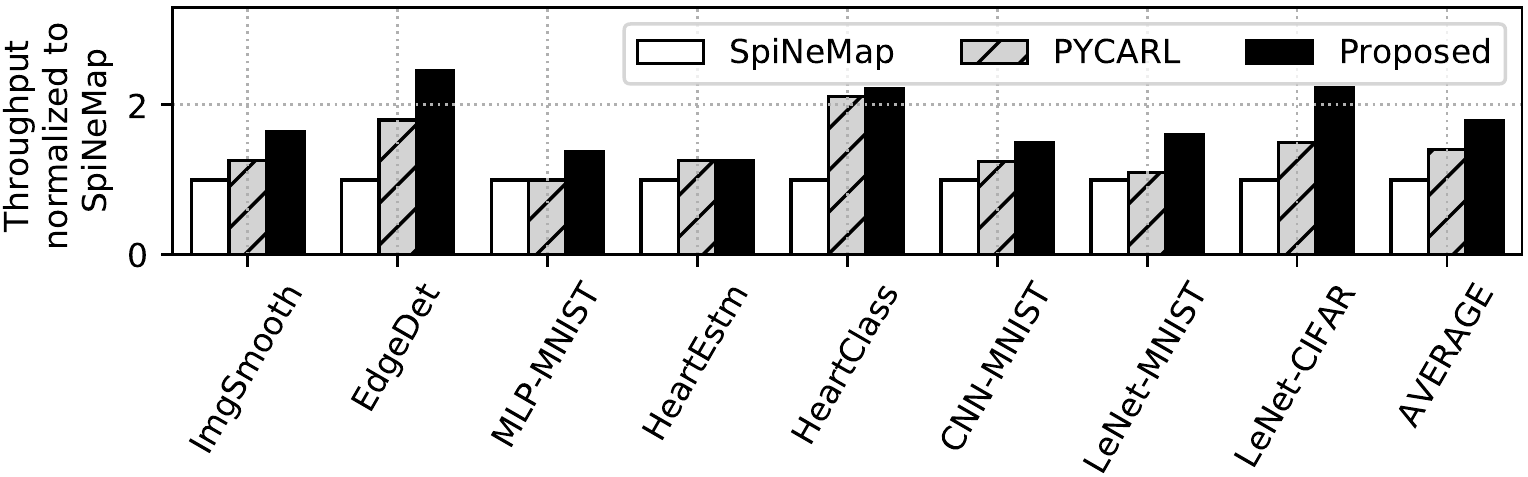}}
	\vspace{-10pt}
	\caption{Throughput, normalized to SpiNeMap.}
	\label{fig:performance}
	\vspace{-10pt}
\end{figure}

Figure~\ref{fig:performance} reports throughput obtained on the DYNAP-SE neuromorphic hardware of each of our application for each of the evaluated techniques normalized to SpiNeMap. We make the following three observations.

First, throughput obtained using SpiNeMap is the lowest among all the evaluated techniques. This is because SpiNeMap places SNN clusters on tiles to minimize the number of inter-tile spikes. Therefore, some tiles need to execute many SNN clusters. As cluster ordering on a tile is not addressed in SpiNeMap, throughput is significantly low. Second, throughput obtained using PYCARL is better than SpiNeMap by an average of 41\%. 
Although PYCARL also orders cluster execution on a tile randomly, throughput of PYCARL is higher than SpiNeMap. 
This is due to PYCARL's strategy to balance the load on each tile, resulting in lower number of clusters mapped per tile than SpiNeMap.
Third, throughput obtained using our approach is the highest (78\% higher than SpiNeMap and 28\% higher than PYCARL). This improvement is due to our static-order schedule, which we analyze and construct at design-time for every tile of the hardware to decide the exact order in which clusters mapped to the same tile need to be executed to improve performance.

\subsection{Cluster Binding}
\begin{figure}[h!]
	\centering
	\vspace{-5pt}
	\centerline{\includegraphics[width=0.99\columnwidth]{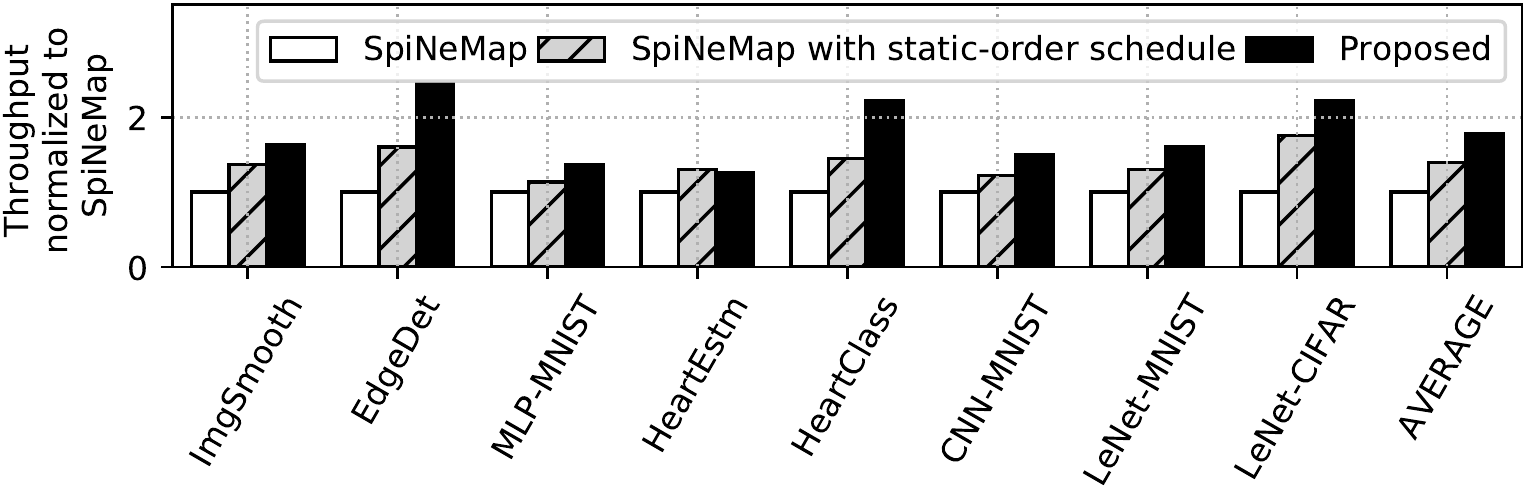}}
	\vspace{-10pt}
	\caption{Throughput, normalized to SpiNeMap.}
	\label{fig:performance_no_mapping}
	\vspace{-10pt}
\end{figure}

We reason that balancing the load on the tiles of a hardware is essential to achieving high throughput. Figure~\ref{fig:performance_no_mapping} reports throughput of each of our application on the DYNAP-SE hardware. We compare our proposed approach against baseline SpiNeMap with random cluster order on each tile and SpiNeMap with static-order schedule on each tile. Throughput results are normalized to SpiNeMap.
We make the following two observations.
First, throughput of SpiNeMap improves by an average of 39\% when static-order scheduling is enabled for each tile of the hardware.
Second, our approach improves throughput further by an average of 27\%. Although the static-order scheduling remains the same, our proposed approach, which balances the load on each tile improves throughput compared to SpiNeMap.

\subsection{Compilation Time}\label{sec:compile_time}
\begin{figure}[h!]
	\centering
	\vspace{-5pt}	\centerline{\includegraphics[width=0.99\columnwidth]{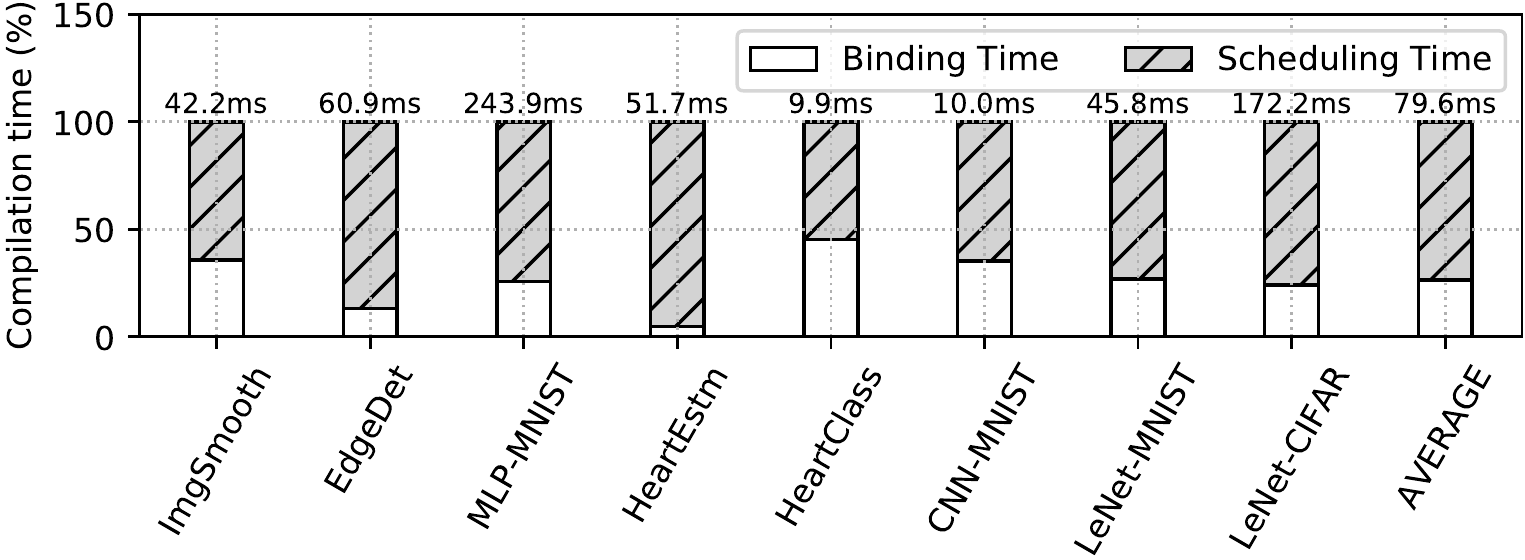}}
	\vspace{-10pt}
	\caption{Fraction of total compile time, distributed into binding and scheduling time.}
	\label{fig:compile_time}
	\vspace{-10pt}
\end{figure}

Figure~\ref{fig:compile_time} reports the fraction of total compilation time of each of our application using our proposed approach for the DYNAP-SE hardware, distributed into time to bind clusters to tiles and the time to construct static-order schedule on each tile. 
The number on each bar reports the absolute time in ms to compile these applications on the DYNAP-SE hardware. We observe that the time consumed to create static-order schedule on each tile is on average 75\% of the total time to compile these applications on the hardware. For some applications such as HeartEstm, the scheduling time is over 95\% of the total time to compile the application.
These results suggest that for run-time use, the schedule construction time needs to be reduced, which justifies our fast self-timed execution based scheduling (see Section~\ref{sec:run_time}).

\subsection{Resource Utilization}
Table~\ref{tab:resource} reports the utilization of hardware resources (tile resources, buffer size, connections, and input and output bandwidth) on the DYNAP-SE neuromorphic hardware for each application. 
The average utilization of hardware resources are 92.5\% for the crossbar IOs on each tile, 9.0\% for buffer space, 42.6\% for connections, and 15\% for input and output tile bandwidth.
Since we perform hardware-aware analysis, resource utilization never exceeds 100\%.

\begin{table}[h!]
	\renewcommand{\arraystretch}{1.0}
	\setlength{\tabcolsep}{3pt}
	\centering
	{\fontsize{7}{10}\selectfont
		\begin{tabular}{|l|c|c|c|c|c|}
			\hline
			\multirow{3}{*}{\textbf{Application}} & \multicolumn{5}{|c|}{\textbf{Utilization (\%)}}\\ \cline{2-6}
			& \multirow{2}{*}{\textbf{Tile}} & \multirow{2}{*}{\textbf{Buffer}} & \multirow{2}{*}{\textbf{Connections}} & 
			\multicolumn{2}{|c|}{\textbf{Bandwidth}}\\ \cline{5-6}
			&&&&\textbf{Input} & \textbf{Output}\\
			\hline
			ImgSmooth & 87.5 & 8.39844 & 37.5 & 17.0898 & 17.0898\\
EdgeDet & 87.5 & 11.2305 & 68.75 & 22.7864 & 22.7865\\
MLP-MNIST & 81.25 & 9.375 & 46.875 & 22.7865 & 22.7864\\
HeartEstm & 96.875 & 9.61914 & 62.5 & 4.70197 & 4.70197\\
HeartClass & 93.75 & 7.91015 & 25 & 9.76564 & 9.76562\\
CNN-MNIST & 100 & 7.42188 & 15.625 & 6.51041 & 6.51041\\
LeNet-MNIST & 93.75 & 8.78906 & 37.5 & 20.3451 & 20.345\\
LeNet-CIFAR & 100 & 9.17967 & 46.875 & 17.0898 & 17.0898\\
			\hline
	\end{tabular}}
	\caption{Resource utilization on DYNAP-SE.}
	\label{tab:resource}
\end{table}

These results illustrate that our approach can also be used for designing  neuromorphic hardware, not only in terms of number of tiles, but all other resources such as buffer space, connections, and input and output bandwidth. 

\subsection{Performance Scalability}
\begin{figure}[h!]
	\centering
	\vspace{-5pt}
	\centerline{\includegraphics[width=0.99\columnwidth]{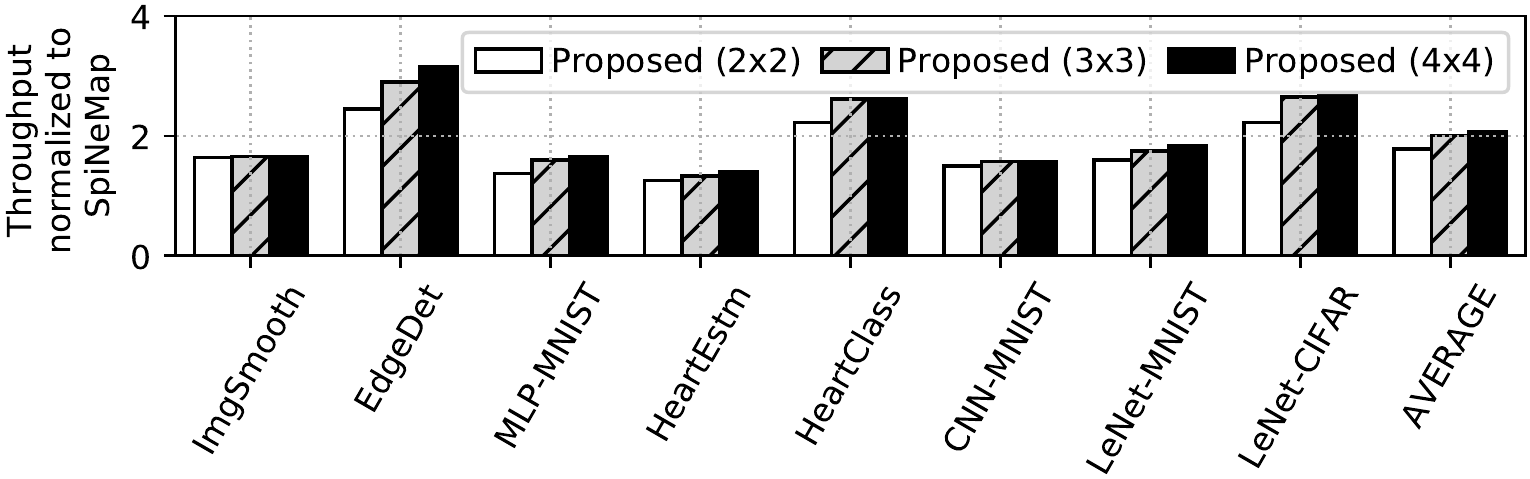}}
	\vspace{-10pt}
	\caption{Throughput, normalized to SpiNeMap.}
	\label{fig:performance_scalability}
\end{figure}

Figure~\ref{fig:performance_scalability} reports throughput of each of our application for our proposed approach normalized to SpiNeMap. 
We compare throughput obtained 
on three hardware models: 4 tiles (baseline), 9 tiles (arranged in a $3 \times 3$), and 16 tiles (arranged in a $4 \times 4$).
We make the following two observations.

First, throughput generally increases with increasing the number of neuromorphic tiles. With 9 and 16 tiles, the average throughput is higher than the baseline configuration by 11\% and 15\%, respectively. This improvement is because with more tiles in the hardware, a tile is shared among fewer clusters, which improves throughput. Second, for applications such as ImgSmooth, four tiles are sufficient to map the application. There is therefore no significant improvement in throughput when the number of tiles in the hardware is increased. For other applications such as EdgeDet, throughput increases with increase in the number of tiles. 

\subsection{Run-time Performance}\label{sec:run_time}
\begin{figure}[h!]
	\centering
	\vspace{-5pt}
	\centerline{\includegraphics[width=0.99\columnwidth]{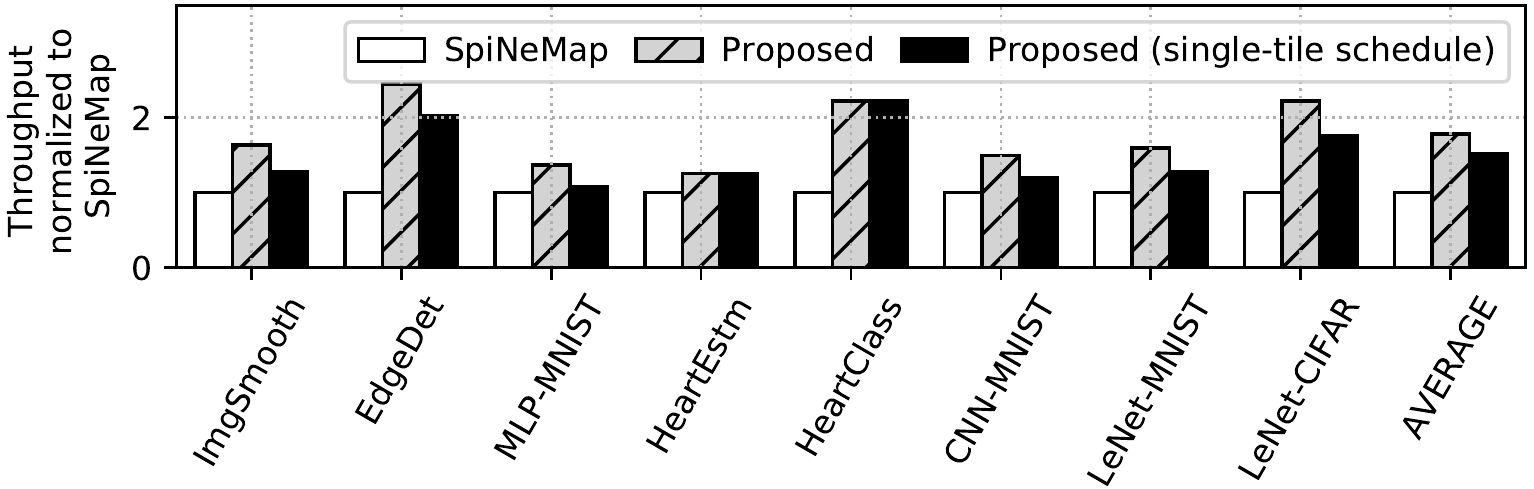}}
	\vspace{-10pt}
	\caption{Throughput, normalized to SpiNeMap.}
	\label{fig:runtime}
	\vspace{-10pt}
\end{figure}

Figure~\ref{fig:runtime} reports throughput of each of our application for our proposed approach normalized to SpiNeMap. 
We compare throughput obtained at design-time where cluster schedules are independently constructed for each tile against throughput obtained at run-time using our proposed single-tile schedule.
We make the following three observations.

First, throughput obtained at run-time from a single-tile static-order schedule is on average 15\% lower than the case when schedules are constructed independently --- that is, by using our design-time analysis method. This verifies Lemma 1. Second, for some application such as HeartEstm and HeratClass, throughput obtained at run time is exactly the same as that obtained at design time. Third, throughput at run time is still higher than SpiNeMap by an average of 51.4\%.

Table~\ref{tab:runtime} compares the compilation time for each application using our approach at design time against that at run time. On average, the run-time approach achieves an average 67.5\% reduction in compilation time. This is due to the reduction of schedule construction overhead using the single-tile static-order schedule along with the self-timed execution approach.

\vspace{-10pt}
\begin{table}[h!]
	\renewcommand{\arraystretch}{1.0}
	\setlength{\tabcolsep}{2pt}
	\centering
	{\fontsize{7}{10}\selectfont
		\begin{tabular}{|l|c|c|l|c|c|}
			\hline
			\multirow{2}{*}{\textbf{Application}} & \multicolumn{2}{c|}{Compilation time (ms)} & \multirow{2}{*}{\textbf{Application}} & \multicolumn{2}{c|}{Compilation time (ms)}\\ \cline{2-3}\cline{5-6}
			& \textbf{Design-time} & \textbf{Run-time} & & \textbf{Design-time} & \textbf{Run-time} \\
			\hline
			ImgSmooth & 42.203 & 17.3685 & EdgeDet	&  60.872	& 12.1011\\
            MLP-CNN	& 243.862	& 77.0835 & HeartEstm	& 51.73	& 6.6567\\
            HeartClass	& 9.939	& 5.0043 & CNN-MNIST	& 10.005	& 4.072\\
            LeNet-MNIST	& 45.829	& 15.1356 & LeNet-CIFAR	& 172.203	& 52.635\\
			\hline
	\end{tabular}}
	\caption{Compilation time at design time vs. at run time.}
	\label{tab:runtime}
\end{table}
\vspace{-24pt}

\subsection{\mc{Accuracy Impact}}
\mc{Table \ref{tab:apps} column 6 reports the model accuracy for each of the evaluated applications obtained on DYNAP-SE. The accuracy is within 5\% of the top accuracy reported in literature. We observe that there is no accuracy loss for applications that are built directly in spiking domain (EdgeDet and ImgSmooth). For all other applications converted from the analog domain, accuracy loss is less than 5\% of the accuracy reported in literature. This loss is attributed to the N2D2 tool.}





\section{Related Works}
\label{sec:related_works}
\subsection{State-of-the-art Neuromorphic Hardware}
\label{sec:neuromorphic_hardware}
In SpiNNaker~\cite{furber2014spinnaker}, each ARM9 core can implement multiple neuron functionality, with the local memory serving as the synaptic storage. 
TrueNorth is a million-neuron digital CMOS chip from IBM~\cite{debole2019truenorth}. The chip has 4,096 tiles, with each tile hosting 12.75 kilobytes of local SRAM memory to store the synapses. 
Loihi is a 128-tile neuromorphic chip from Intel, with each tile having 1,024 spiking neurons and 2 Mb of SRAM to store synapses\cite{davies2018loihi}. 
There are also many other neuromorphic chips such as Neurogrid \cite{benjamin2014neurogrid}, BrainScaleS \cite{schemmel2012live}, Braindrop \cite{neckar2018braindrop}, and ODIN \cite{frenkel20180}. These architectures are similar to 
DYNAP-SE~\cite{Moradi_etal18}, which we evaluate.

\subsection{Mapping SNNs to Neuromorphic Hardware}
Corelet is a proprietary tool from IBM to map SNNs to TrueNorth~\cite{amir2013cognitive}. PACMAN is used to map SNNs to SpiNNaker~\cite{galluppi2015framework}. Beyond these hardware-specific tools, there are also general-purpose ones. For instance, PyNN~\cite{davison2009pynn} is used to map SNNs on Loihi, BrainScaleS, SpiNNaker, and Neurogrid by balancing the load on each tile. The PSO-based technique developed in Das et al. is used to map SNNs to a hardware, reducing the energy consumption between tiles\cite{das2018mapping}. SpiNeMap reduces the communication between tiles~\cite{balaji2019mapping}. PYCARL is proposed to perform hardware-software co-simulation of SNNs \cite{balaji20pycarl}. We compare our approach against PYCARL and SpiNeMap, and found it to perform significantly better.

There are also other approaches that use a single large crossbar to map SNNs~\cite{ankit2018neuromorphic,zhang2018neuromorphic,xia2019memristive,lee2019system,wijesinghe2018all,wen2015eda}.

\subsection{Non-volatile Memory}
Recently, NVMs are used to lower the energy consumption of von-Neumann computing \cite{songISMMa,songISMMb,song2019enabling} and neuromorphic computing. To this end, 
Ramasubramanian et al. use STT-MRAM~\cite{ramasubramanian2014spindle}, Burr et al. use PCM~\cite{burr2017neuromorphic}, and Mallik et al. use OxRAM~\cite{mallik2017design} to design neuromorphic tiles.

\subsection{Similar Concept in Related Domain}
SDFGs are widely used for predictable mapping of applications to multiprocessor systems. Numerous approaches to throughput analysis of SDFGs have been previously proposed~\cite{stuijk2006exploring,stuijk2007multiprocessor,damavandpeyma2012modeling,zhu2012static}. Bonfietti et al. evaluated mappings of SDFG to multiprocessor system, maximizing the throughput~\cite{bonfietti2013maximum}. Stemmer et al. propose to use probabilistic analysis to allocate and schedule SDFGs on multiprocessor systems~\cite{stemmer2020towards}. Das et al. evaluated the fault-tolerant mapping of SDFGs to multiprocessor systems~\cite{DBLP:journals/tecs/DasKV14,das2015reliability,das2014communication}. Recently, SDFG-based analysis is also proposed for analyzing machine learning applications~\cite{balaji2019framework,das2018dataflow,hong2017hierarchical,chen2017using,bacis2017pipelined}. However, none of these approaches address application analysis with limited hardware resources, both at design-time and at run-time.

\section{Conclusions}
\label{sec:conclusions}

We introduce an approach for predictable compilation of SNN-based applications on state-of-the-art neuromorphic hardware. Prior works have only addressed design-time mapping, considering unlimited resources in the underlying hardware. These approaches present significant limitations when used to compile and execute machine learning applications on a resource-constrained hardware. Our approach makes three contributions. First, we propose a technique to generate neuron and synapse clusters, where each cluster can fit on to the resources of a tile of the hardware. Second, we exploit the rich semantics of SDFG to model SNN clusters and analyze performance on a resource-constrained hardware. Finally, we propose a scheduling approach based on self-timed execution to reduce the time to compile and admit an application to a hardware at run-time, adjusting to dynamic resource availability.
We conducted experiments with standard SNN-based applications and demonstrate a significant increase in performance 
over current practices.

\section{Acknowledgments}
This work is supported by 1) the National Science Foundation Award CCF-1937419 (RTML: Small: Design of System Software to Facilitate Real-Time Neuromorphic Computing) and 2) the National Science Foundation Faculty Early Career Development Award CCF-1942697 (CAREER: Facilitating Dependable Neuromorphic Computing: Vision, Architecture, and Impact on Programmability).

\bibliographystyle{ACM-Reference-Format}
\bibliography{snn2sdfg}

\end{document}